\documentclass[aps,pre,twocolumn,superscriptaddress,amssymb]{revtex4-2}
\usepackage[dvipdfmx]{graphicx}
\usepackage[dvipdfmx]{color}
\usepackage{physics}
\usepackage{color}
\usepackage{listings}
\usepackage{bm}
\usepackage{amsmath}
\usepackage{amsfonts}
\usepackage{appendix}
\usepackage{mathtools}
\usepackage{comment}

\newcommand{\argmax}{\mathop{\rm arg~max}\limits}
\newcommand{\argmin}{\mathop{\rm arg~min}\limits}

\begin{document}

\title{Phase transition in compressed sensing with horseshoe prior}
\author{Yasushi Nagano}
\affiliation{
 Graduate School of Arts and Sciences,
 The University of Tokyo,
 Komaba, Meguro-ku, Tokyo 153-8902, Japan
}

\author{Koji Hukushima}
\affiliation{
 Graduate School of Arts and Sciences,
 The University of Tokyo,
 Komaba, Meguro-ku, Tokyo 153-8902, Japan
}
\affiliation{
Komaba Institute for Science, The
University of Tokyo, 3-8-1 Komaba, Meguro-ku, Tokyo 153-8902, Japan
}
\begin{abstract}
    In Bayesian statistics, horseshoe prior has attracted increasing attention as an approach to the sparse estimation. The estimation accuracy of compressed sensing with the horseshoe prior is evaluated by statistical mechanical method. It is found that there exists a phase transition in signal recoverability in the plane of the number of observations and the number of nonzero signals and that  the recoverability phase is more extended than that using the well-known $l_1$ norm regularization.
\end{abstract}
\date{\today}

\maketitle

\section{Introduction}
Compressed sensing (CS) is a generic term for a set of methods for estimating high-dimensional parameters from a smaller number of observations; it is a type of sparse modeling\cite{donoho2006compressed,candes2008introduction}. In observations with high measurement costs, such as magnetic resonance imaging, undersampling is desirable for fast observations, and it is important to accurately recover high-dimensional image data from a small number of observations.
CS allows such estimation by assuming sparsity, that is, assuming that the data can be represented by a small number of intrinsic low-dimensional parameters, and by obtaining them with a high degree of accuracy\cite{donoho2009observed}.

CS is often formulated as a problem of optimizing the regularization functions of the parameters to be estimated. The low-dimensional parameters are obtained by minimizing the regularization function under the constraint that the parameters are consistent with the observations, especially for noiseless CS. Therefore, the regularization used significantly affects the accuracy of CS estimation.

The $ l_1$ norm of parameters is commonly used as a regularization function. Regression using $ l_1$ norm regularization is guaranteed to find sparse solutions, and because the optimization problem is a convex optimization problem, the solution can be obtained with computational efficiency \cite{tibshirani1996regression,tibshirani1997lasso,chen2001atomic,beck2009fast}. Although $ l_1$ norm regularization is very commonly used because of these desirable properties, several drawbacks have been noted. In particular, the residual bias in the estimation of nonzero components and the high false-positive rate (FPR) are serious issues in practice. 

Many efforts have been made to solve the problems of $l_1$ norm regularization. In recent years, regression with hierarchical Bayes models has attracted much attention from researchers in statistics.
In Bayesian statistics, CS is equivalent to maximum a posteriori (MAP) estimation with a prior distribution corresponding to regularization. Therefore, finding a good regularization is equivalent to finding a good prior distribution.

The prior distribution corresponding to $l_1$ norm regularization is the Laplace distribution. Note that the Laplace distribution is obtained by mixing the variance of each element of the Gaussian distribution, whose variance-covariance matrix is a diagonal matrix, according to the exponential distribution\cite{park2008bayesian}. 
Sparse regression with a hierarchy of variance parameters has been proposed, among which the horseshoe (HS) prior is considered the most promising \cite{carvalho2009handling,carvalho2010horseshoe,bhadra2019lasso}. 

In the HS prior, the standard deviation of the Gaussian distribution is expressed as the product of the global scale hyperparameter common to all entries and the local scale parameters
assigned to each entry; its prior distribution is a half-Cauchy distribution. 

This prior distribution is proposed because the shrinkage parameter of $ l_2$ norm regularization corresponds to the Gaussian distribution and is expected to yield sparse solutions as in the $l_1$ norm regularization.  In addition, it overcomes some drawbacks of $ l_1$ norm regularization that need to be addressed, demonstrating that the bias in estimating nonzero entries is asymptotically zero and that the  bounds of the Bayes risk are more stringent than those in $ l_1$ norm regularization in some cases\cite{van2014horseshoe}. Theoretically, the validity of using the half-Cauchy distribution as the variance parameter in the hierarchical prior and the advantage of having both global and local scale parameters have been noted\cite{gelman2006prior,polson2012half,van2016conditions,ghosh2016asymptotic}. Furthermore, the ability to use an efficient algorithm with Gibbs sampling to sample from the posterior distribution is among the practical advantages\cite{makalic2015simple,bhattacharya2016fast}. 

Statistical physics approaches, on the other hand,  have also been used extensively to study CS. For example, the existence of phase transitions in CS has been analyzed theoretically using $l_p$ norm regularization with $p=0, 1, 2$ to derive exact phase boundaries\cite{kabashima2009typical,ganguli2010statistical}. Furthermore, the convergence of the approximate message-passing algorithm\cite{donoho2009message,donoho2010message_2} that approximates belief propagation has been analyzed theoretically\cite{caltagirone2014convergence}, and CS and the derivation of its algorithms have been analysed by Bayesian inference using prior distributions that explicitly include  sparsity\cite{krzakala2012statistical,krzakala2012probabilistic}. These statistical physics analyses were performed assuming that the number of observations is infinite and the ratio of the number of parameters to that of observations is a positive constant. In contrast, previous theoretical studies in statistics of  HS prior distributions have analyzed the asymptotic behavior for an infinite number of observations relative to the number of parameters. Since the statistics of small data is an important subject in Bayesian statistics, it is also important to understand the behavior when the number of observations is close to the number of parameters. 

In this paper, we discuss the typical recoverability of CS for random observations using  HS regularization with a fixed global scale parameter to quantitatively confirm the usefulness of the half-Cauchy distribution as a prior distribution of variance in a hierarchical prior. The discussion consists mainly of an analysis of the average free entropy of the system with respect to the observations using the replica method of statistical physics. 
We consider the limit of an infinite number of observations with the ratio between the number of observations and that of parameters being $\order{1}$, 
which is commonly used in statistical physics, as mentioned above.  For HS regularization, we confirm that the solution assuming replica symmetry (RS) gives exactly the recoverable phase. We also investigate the recoverability limit on the plane of the observed number density and the nonzero signal number density and find that HS regularization has a wider recoverable region than $ l_1$ norm regularization. HS regularization also has a lower FPR, indicating that the use of the half-Cauchy distribution for the prior distribution is effective even when only the local scale parameters are considered.

The rest of this paper is organized as follows. In Sec.~\ref{sec:2}, we introduce CS with the HS prior and give the formulation in terms of statistical mechanics. In Sec.~\ref{sec:3}, we present theoretical results such as the phase boundaries and FPR obtained by the replica method. In Sec.~\ref{sec:4}, we demonstrate and discuss numerical results, which are in agreement with the result of our theory. Section~\ref{sec:5} presents a summary and discussion. The Appendices A, B, and C present the details of the analysis. 

\section{Problem settings and Formulation}
\label{sec:2}
\subsection{Compressed sensing with horseshoe prior}

CS is a method of inferring $N$-dimensional parameters with $M$ measurements, where $M$ is smaller than $N$. In general, this is not possible. However, if the true parameter $\bm w_0$ is expected to be sparse and the number of nonzero elements $P$ is smaller than $M$, successful signal recovery may be possible. In noiseless CS considered in this study, the true parameter $\bm w_0$ is inferred from the observations $\mathcal{D} = \{\bm y, X\}$, which consist of the measurements $\bm y$ and design matrix $X$ as 
\begin{equation}
    \bm y = X\bm w_0,~\bm y\in\mathbb{R}^M,~X\in\mathbb{R}^{M\cross N},~\bm w_0\in\mathbb{R}^N. 
    \label{eqn:linearR}
\end{equation}
Noiseless CS can be defined as an optimization problem in which the normalization term (or Hamiltonian) $\mathcal{H}(\bm w)$ is minimized under the constraint $\bm y = X\bm w$:
\begin{equation}\label{eqn:optimization}
    \bm w^* = \argmin_{\bm w} \mathcal{H}(\bm w)~\mbox{s.t.}~\bm y = X\bm w,~\bm w\in\mathbb{R}^N. 
\end{equation}
The $ l_1$ norm regularization is well-known as a typical regularization term, whose normalization term is given by 
\begin{equation}
\mathcal{H}(\bm w)=\lambda|\bm w|_1 = \lambda\sum^N_{i=1}|w_i|
\end{equation}
for $\lambda > 0$.

In Bayesian statistics, CS can be described as MAP estimation with a sparse prior distribution $p(\bm w)$: 
\begin{equation}
\begin{split}
    \bm w^* &= \argmax_{\bm w} p(\bm w|\mathcal{D}),\\
    &=\argmax_{\bm w} p(\mathcal{D}|\bm w)p(\bm w),\\
    &=\lim_{\gamma\rightarrow\infty}\argmax_{\bm w} e^{-\frac{\gamma}{2}|\bm y-X\bm w|^2}e^{-\mathcal{H}(\bm w)},
\end{split}
\end{equation}
where $\gamma$ is a positive constant. 
Clearly, the $l_1$ norm regularization corresponds to MAP estimation using the  Laplace distribution for the prior distribution. 
For the HS prior, the prior distribution of $w_i$ is Gaussian, but its variance is a random variable with a half-Cauchy distribution $C^+(0,1)$ as the prior distribution: 
\begin{equation}
    w_i\sim\mathcal{N}(0,\tau^2\lambda_i^2),
\end{equation}
\begin{equation}
    \lambda_i\sim C^+(0,1) = \frac{2}{\pi}\frac{1}{1+\lambda^2},
    \label{eqn:Cauchy}
\end{equation}
where $\lambda>0$, and $\tau$ is a hyperparameter corresponding to global scale parameter. This hyperparameter can be determined using a value that seem reasonable or by cross-validation or a fully Bayesian approach. Here, we focus on the case of $\tau=1$ because we are interested in the validity of the prior of the local scale parameters. From the prior distribution, the regularization is given by $-\log p(\bm w)$, and the term in HS prior can be written explicitly as 
\begin{equation}\label{eqn:HofHS}
\begin{split}
    \mathcal{H}(\bm w) &= \sum^N_{i=1} \mathcal{H}(w_i),\\
    &= \sum^N_{i=1} -\ln\int\dd{\lambda_i}C^+(\lambda_i|0,1)\mathcal{N}(w_i|0,1). 
\end{split}
\end{equation}

\begin{figure}
    \centering
    \includegraphics[width=\linewidth]{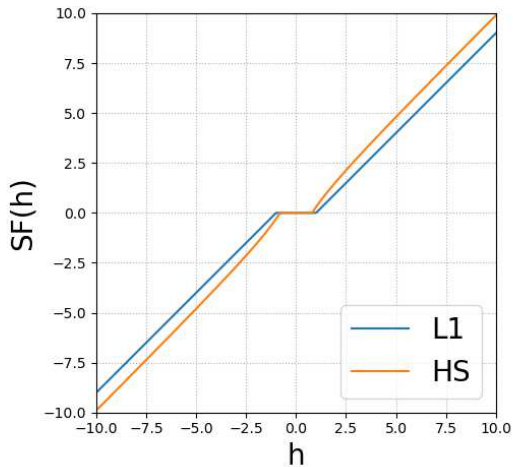}
    \caption{Score function of $ l_1$ norm regularization with $\lambda=1$ (L1)  and HS regularization with $\tau=1$ (HS). In both cases, the output is zero for small inputs. For HS regularization, the score function asymptotically approaches $y=x$ for sufficiently large inputs, whereas for $l_1$  regularization, it is always $y\neq x$.}
    \label{fig:scorefunc}
\end{figure}

The score function, which intuitively characterizes sparse linear regression, is defined as 
\begin{equation}
    \mbox{SF}(h) = \argmax_w -\frac{1}{2}|w|^2 + hw - \mathcal{H}(w). 
\end{equation}
As shown in Fig. \ref{fig:scorefunc}, the output of the score function is asymptotically equal to the input for large inputs in HS regularization. However, this is not asymptotically the case for $l_1$ regularization, which indicates the existence of an estimation bias. 

\subsection{Formulation}
In the setting discussed in this paper, one infers the true parameter $\bm w_0$ given the design matrix $X$ and observation $\bm y = X\bm w_0$. 
In the following analysis of Gaussian random problems, the design matrix and true parameters are assumed to be generated following i.i.d. as 
\begin{equation}
    X_{ij}\sim\mathcal{N}\left(0,\frac{1}{N}\right),
    \label{eqn:X}
\end{equation}
and 
\begin{equation}
    w_{0,i}\sim\rho\mathcal{N}(0,1)+(1-\rho)\delta(w_{0,i}),
    \label{eqn:w}
\end{equation}
respectively, where $\rho$ is the fraction of nonzero elements in the true parameters. 
From a statistical mechanics viewpoint,  the nature of CS is revealed by the behavior of the free-entropy density $\phi(\beta;X,\bm w_0)$ as $\beta\rightarrow\infty$, which is defined as 
\begin{equation}
    \phi(\beta;X,\bm w_0) = \frac{1}{N}\ln Z(\beta;X,\bm w_0), 
\end{equation}
where the partition function is expressed using the regularization term ${\cal H}(\omega)$ as  
\begin{equation}
    Z(\beta;X,\bm w_0) = \int \dd{\bm w}\delta(X\bm w_0-X\bm w)\exp(-\beta\mathcal{H}(\bm w)). 
\end{equation}
We here introduce the hierarchical prior as $\beta\rightarrow\infty$ as 
\begin{equation}\label{eqn:hierarchical_prior}
    \exp(-\beta\mathcal{H}(\bm w)) \propto \int^\infty_0\prod^N_{i=1}\dd{\lambda_i}\pi^\beta(\lambda_i)~\mathcal{N}\qty(w_i|0,\beta^{-1}\lambda_i^2), 
\end{equation}
where $\pi(\lambda)$ can be taken to be a prior distribution one level up the hierarchy, in which the variance of the Gaussian distributions is mixed. 
For instance, in the HS prior mentioned in the previous section, $\pi(\lambda)$ is the half-Cauchy distribution given by Eq.~(\ref{eqn:Cauchy}). 

According to statistical mechanics, especially the spin-glass theory, under the limit $N\rightarrow\infty$ with $\alpha\equiv M/N$ fixed, the free-entropy density is averaged with respect to $X$ and $\bm w_0$ assuming the self-averaging property described as 
\begin{equation}
    \phi(\beta;\alpha,\rho) = \mathbb{E}_{X,\bm w_0}\qty[\phi(\beta;X,\bm w_0)], 
\end{equation}
where $\mathbb{E}_{X,\bm w_0}\qty[\cdots]$ represents the expectation value for the distributions  of $X$ and $\bm w_0$ in Eqs.~(\ref{eqn:X}) and (\ref{eqn:w}), respectively. 
The right-hand side of this equation can be evaluated using the replica method as 
\begin{equation}
    \phi(\beta;\alpha,\rho) = \frac{1}{N}\lim_{n\rightarrow0}\pdv{}{n}\mathbb{E}_{X,\bm w_0}\qty[Z^n]. 
\end{equation}

\section{Analytical results}
\label{sec:3}
While details of the calculations are given in Appendix~\ref{sec:RS_free_energy}, 
the calculations assuming RS yield the RS free-energy density of the hierarchical prior in the limit  of $\beta\rightarrow\infty$ as 
\begin{align}
     \frac{1}{\beta}\phi(\beta;\alpha,\rho) &= \max_{\hat Q,\hat\chi,\hat m,q,\chi,m}\frac{1}{2} q\hat Q - \frac{1}{2}\chi\hat\chi - m\hat m\nonumber\\
     & + \alpha\Psi_{RS}(q,\chi,m;\rho) + \hat\Psi_{RS}(\hat Q,\hat\chi,\hat m), 
     \label{eqn:RS_FED}
\end{align}
where
\begin{equation}
    \Psi_{RS}(q,\chi,m;\rho) = -\frac{\rho-2m+q}{2\chi},
\end{equation}
and 
\begin{align}\label{eqn:hat_psi_rs}
     \hat\Psi_{RS}(\hat Q,\hat\chi,\hat m) &= \mathbb{E}_{w_0,\xi}\left[\frac{1}{\beta}\ln\int^\infty_0\dd{\lambda} \right. \nonumber \\
     & \left.\exp(\beta\qty(\frac{(w_0\hat m+\sqrt{\hat\chi}\xi)^2}{2(\hat Q+\lambda^{-2})}+\ln\pi(\lambda)))\right],
\end{align}
with $\xi\sim\mathcal{N}(0,1)$. 

In Eq.~(\ref{eqn:RS_FED}), the order parameters $q$, $\chi$, and $m$ and the conjugate parameters $\hat{Q}$, $\hat{\chi}$, and $\hat{m}$ are the variables determined to maximize the right-hand side. Their maximization conditions are called the saddle point equations, which are derived from the partial derivative of the free-energy density with respect to these parameters as follows:
\begin{equation}
 \begin{split}
     q &= \mathbb{E}_{w_0,\xi}\qty[\left\langle\qty(\frac{w_0\hat m+\sqrt{\hat\chi}\xi}{\hat Q+\lambda^{-2}})^2\right\rangle],\\
    m &= \mathbb{E}_{w_0,\xi}\qty[\left\langle\frac{w_0^{2}\hat m+w_0\sqrt{\hat\chi}\xi}{\hat Q+\lambda^{-2}}\right\rangle],\\
    \chi &= \mathbb{E}_{w_0,\xi}\qty[\left\langle\frac{\xi^2+\frac{w_0\hat m}{\sqrt{\hat\chi}}\xi}{\hat Q+\lambda^{-2}}\right\rangle],\\
    \hat Q &= \hat m = \alpha\frac{1}{\chi},\\
    \hat\chi &= \alpha\frac{\rho-2m+q}{\chi^2},
\end{split}
\label{eqn:spe}
\end{equation}
where $\langle\cdots\rangle$ is defined as 
\begin{equation}
    \left\langle f(\lambda) \right\rangle = f(\lambda^*), 
\end{equation}
with
\begin{equation}\label{decide_variance}
    \lambda^* = \argmax_\lambda ~\ln \pi(\lambda)+ \frac{(w_0\hat m+\sqrt{\hat\chi}\xi)^2}{2(\hat Q+\lambda^{-2})}.
\end{equation}
 When the true signal is successfully reconstructed, the limit $\chi\rightarrow0$ is a necessary condition for the inference to be well-concentrated as $\beta\rightarrow\infty$. In this limit, among the conjugate parameters, only $\hat\chi$ can have a nontrivial solution of the saddle point equations. Then, by eliminating other order parameters, the self-consistent equation of $\hat\chi$ is obtained as 
\begin{align}
     \hat\chi &= \frac{1}{\alpha}\left(\rho\qty(\hat\chi+\frac{5}{2}-\frac{1}{2}e^4\mbox{erfc}(2)-\frac{2}{\sqrt{\pi }})\right. \nonumber \\
     & \left.{\ \ }+2(1-\rho)\int_{\xi>\sqrt{\frac{2}{\hat\chi}}}D\xi\qty(\sqrt{\hat\chi}\xi-\sqrt{2})^2\right),
\end{align}
where $D\xi=\exp\left(-\xi^2/2\right)d\xi/\sqrt{2\pi}$.  See Appendix~\ref{Sec:B} for details. 

\begin{figure}
    \centering
    \includegraphics[width=\linewidth]{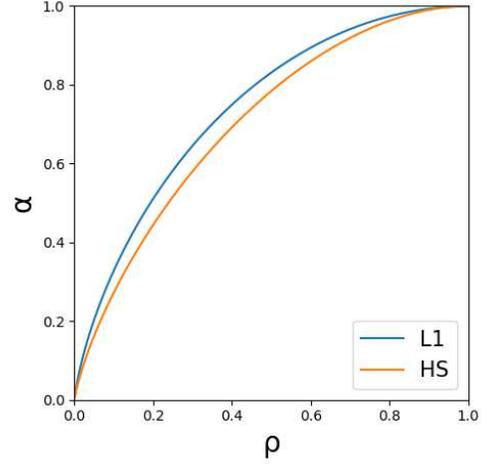}
    \caption{Phase boundaries of signal recoverability in CS for $ l_1$ norm regularization (L1) and HS regularization (HS). Complete recovery is possible in the region above the phase boundary. }
    \label{fig:PT}
\end{figure}

The condition for this equation to have a stable fixed point is obtained from the linear stability of that fixed point as 
\begin{equation}\label{linear_stability_condition}
    \alpha > \rho+2(1-\rho)\Phi\qty(\sqrt{\frac{2}{\hat\chi}}), 
\end{equation}
where $\Phi(x) = \int^\infty_x D\xi$. 
Figure~\ref{fig:PT} shows this condition obtained numerically, together with the condition for $ l_1$ norm regularization. This condition represents the signal recovery bound and is the HS regularization version of the Donoho-Tanner transition line in $l_1$ norm regularization. HS regularization enhances the signal recovery phase slightly but significantly compared to that of $l_1$ norm regularization, indicating that CS by HS regularization is superior to that by $ l_1$ norm regularization, independent of the value of $\rho$. 

The FPRs of the two regularizations differ remarkably.  
The estimate of $w_i$ will not necessarily be zero if $\lambda^*$, as defined in Eq.~(\ref{decide_variance}), does not equal zero. Consequently, the FPR is derived as 
\begin{equation}
    \mathrm{FPR} = 2\Phi\qty(\sqrt{\frac{2}{\hat\chi}}).
\end{equation}

\begin{figure}
    \centering
    \includegraphics[width=\linewidth]{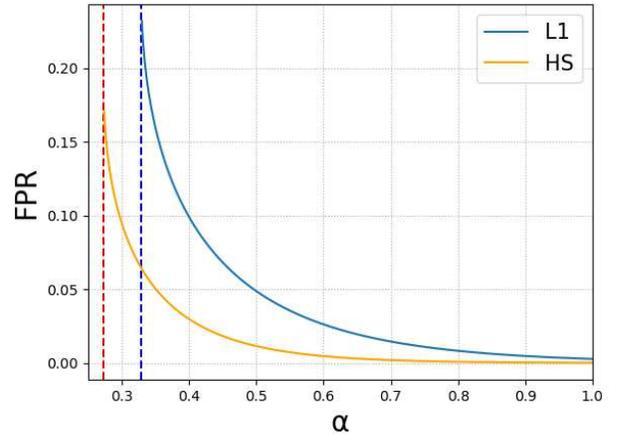}
    \caption{False positive ratio (FPR) as a function of $\alpha$ in the signal recovery phase at $\rho=0.1$ for $ l_1$ regularization (L1) and HS regularization (HS). Two vertical dot lines represent the signal recover bound for $ l_1$ and HS, respectively. }
    \label{fig:fpr}
\end{figure}

Figure~\ref{fig:fpr} shows the FPR in the signal recovery phase evaluated numerically for $\rho=0.1$ using $l_1$ norm regularization and HS regularization. The signal recovery phase is expanded in HS regularization compared to that in $l_1$ norm regularization and the FPR is also significantly improved in the entire parameter region. 

\section{Numerical experiments}
\label{sec:4}
Numerical experiments for finite-size instances were performed to examine the theoretical results obtained using the replica method in the limit $N\rightarrow\infty$ in the previous section. Given the fraction $\rho$ of nonzero elements in the $N$-dimensional true signal $\bm w_0$, the design matrix $X$ is generated according to Eq.~(\ref{eqn:X}), where the number of observations is $M$, and $\bm w_0$ is set according to Eq.~(\ref{eqn:w}) as 
\begin{equation}
    w_{0,i} = \left\{
    \begin{array}{ll}
       \mathcal{N}(0,1)  & \mbox{for}~ i=1,\cdots,N\rho, \\
         0 & \mbox{for}~ i=N\rho+1,\cdots,N.  
    \end{array}
    \right.
\end{equation}
From the generated $X$ and $\bm y = X\bm w_0$, the signal $\bm w$ is inferred from the optimization problem of Eq.~(\ref{eqn:optimization}) with the Hamiltonian in Eq.~(\ref{eqn:HofHS}). For the optimization, we adopt the approximate message-passing algorithm defined by the following iterative equations\cite{donoho2009message}: 
\begin{equation}
\begin{split}
    &\bm z^t = \bm y-X\bm w^t + \bm z^{t-1}\sum^N_{i=1}\eta'(h^{t-1}_i;\gamma^{t-1}),\\
    &\gamma^{t} = \frac{\gamma^{t-1}}{\eta'(X^\top\bm z^{t-1} + \alpha w^t;\gamma^{t-1})}, \\
    &\bm h^t = X^\top\bm z^t + \alpha w^t,\\
    &w^{t+1}_i = \eta(h^t_i;\gamma^t),
\end{split}
\end{equation}
where
\begin{equation}
    \eta(h;\gamma) = \frac{\gamma h}{\gamma\alpha + (\lambda^*)^{-2}},
\end{equation}
and
\begin{equation}
\begin{split}
    &(\lambda^*)^2 \\
    &= \frac{((\gamma h)^2-4\gamma\alpha)+\sqrt{((\gamma h)^2-4\gamma\alpha)^2+8(\gamma\alpha)^2((\gamma h)^2-2)}}{4(\gamma\alpha)^2}.
\end{split}
\end{equation}
In practice, the convergence could be improved by including a damping factor in the update of $w$. 

\begin{figure}
    \centering
    \includegraphics[width=\linewidth]{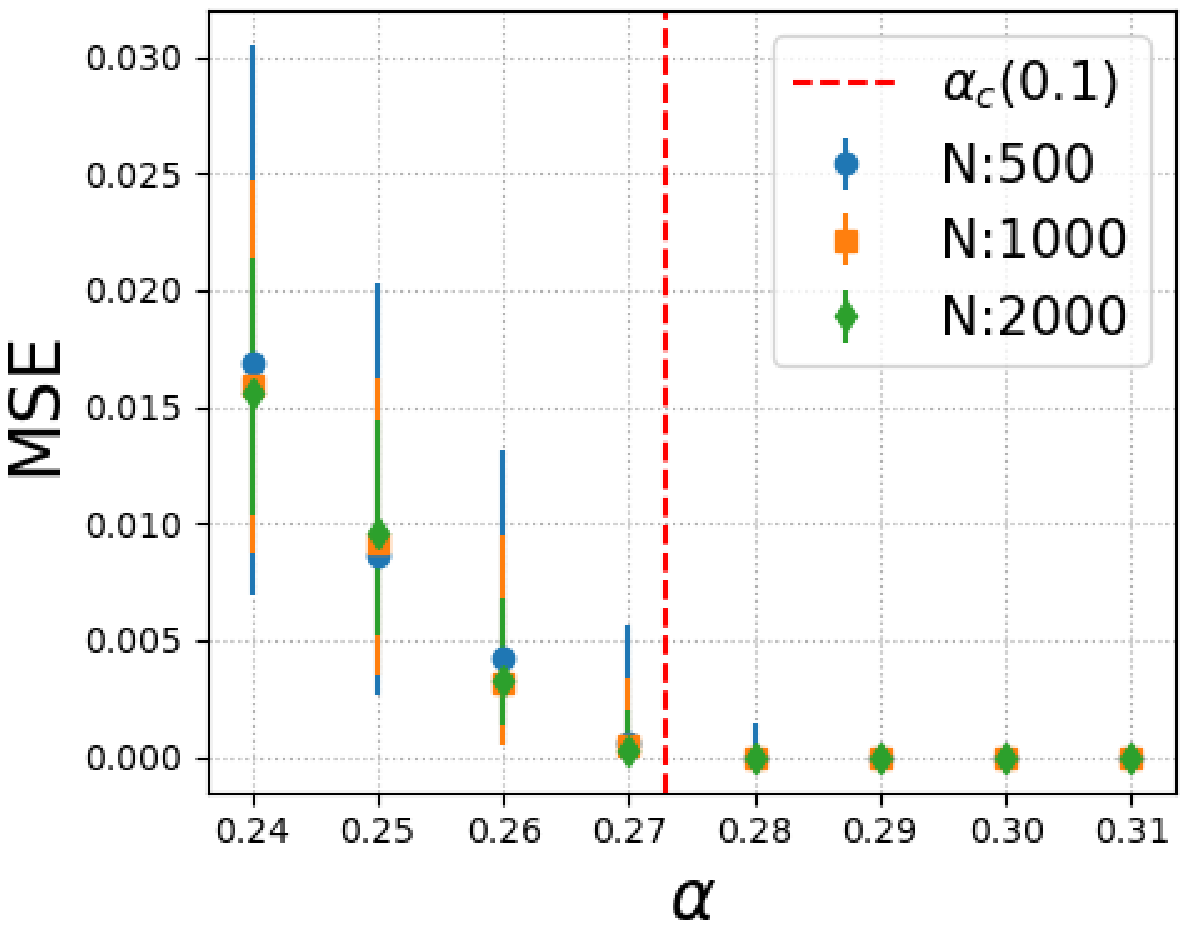}
    \caption{MSE as a function of the observed number density $\alpha$ for CS with $N=500, 1000, 2000$ and $\rho=0.1$ near the phase transition point. For each $\alpha$, the median MSE for $10^3$ independent instances are plotted. The bars represent the quartile ranges of $10^3$ instances. The vertical dotted line represents the transition point obtained using the replica theory,  $\alpha_c(0.1)=0.272(8)$. 
    }
    \label{fig:pt_detail_AMP}
    \includegraphics[width=\linewidth]{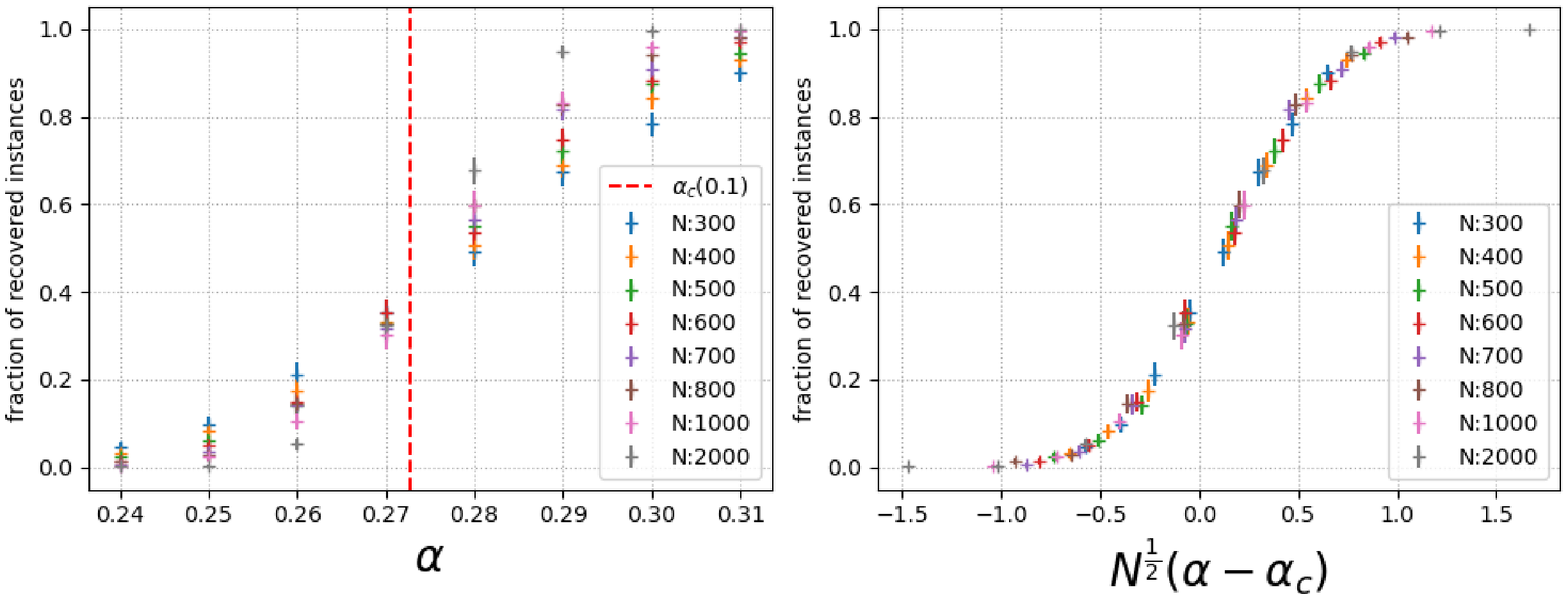}
    \caption{Left: fraction of recovered instances as a function of $\alpha$ of CS with several sizes and $\rho=0.1$. Right: finite-size scaling plot of the fraction of recovered instances of CS with a scaling exponent $\beta = 2.0$ in Eq.~(\ref{eqn:FSS}). The scaling analysis assumes the expected transition point $\alpha_c(0.1)=0.272(8)$ from the RS solution. }
    \label{fig:FSS_AMP}
\end{figure}

We examine in detail the region near the phase transition when $\alpha$ is varied with a fixed $\rho=0.1$ for various system sizes $N$. Figure~\ref{fig:pt_detail_AMP} shows the median MSE for $10^3$ independent instances as a function of $\alpha$. In finite-size experiments, the true signal may be successfully recovered in the sense of $\mathrm{MSE}<10^{-14}$ in some instances when $\alpha$ is smaller than the transition point $\alpha_c$, but such instances are rare. By contrast, for $\alpha>\alpha_c$, many instances are recovered, and the median value is almost zero. A finite-size scaling analysis was performed for more quantitative verification of the transition point obtained by the replica theory. Near the transition point $\alpha_c$, we assumed that a finite-size scaling form of the fraction of recovered instances is expressed as 
\begin{equation}
   \mathrm{Recovered}(N,\alpha) = f\qty(N^{\frac{1}{\beta}}(\alpha-\alpha_c)), 
   \label{eqn:FSS}
\end{equation}
where $f$ is a scaling function and $\beta$ is a scaling exponent. We evaluated the fraction of recovered instances over $10^3$ instances for various $N$ and $\alpha$. Figure~\ref{fig:FSS_AMP} presents the finite-size scaling plot of the fraction of recovered instances with the scaling exponent $\beta=2.0$, which is confirmed to work well assuming the transition point from the replica analysis. 

\section{Summary and Discussion}
\label{sec:5}
As part of the nature of the regression using the HS prior, a statistical-mechanics approach has revealed the phase boundary of the recoverable phase achieved by the HS regularization in CS.  Furthermore, we theoretically derived the typical behavior of the FPR, which is important in sparse linear regression, and found it to be significantly improved over that of the commonly used $ l_1$ regularization. 
Our analysis here assumed the RS ansatz, and its validity must be examined. For the HS prior, the de~Almeida-Thouless stability of the RS solution as $\chi\rightarrow0$ yields the equivalent condition to the linear stability of $\hat\chi$ in Eq.~(\ref{linear_stability_condition}). Thus, the RS ansatz is a reasonable assumption in the recoverable phase. Details of the calculations are given in Appendix~\ref{Sec:C}.
Note that these results are based on a typical evaluation, and the recoverable phase can be considered to have only a negligible probability of failing as $N\rightarrow\infty$, which is different from the worst-case evaluation. In addition, although the calculations using the replica method are in good agreement with numerical experiments, they have not been rigorously proven\cite{talagrand2003spin}.

The HS prior was originally noticed in the context of Bayesian inference, where the combination of global and local scale parameters is thought to play an important role in achieving good performance. However, CS for $\beta\rightarrow\infty$ discussed here is a MAP estimation, that is, a point estimation, and does not fully exploit the characteristics of the Bayesian posterior distribution. Moreover, this study focused on the choice of prior for the local scale parameters, ignoring the effects of the global scale parameter. Thus, this study provides a theoretical analysis of only certain partial features of the HS prior. Nevertheless, the HS prior is shown to be superior to the Laplace prior. This result indicates that the half-Cauchy distribution is preferable as the prior distribution of the local scale parameters.

Our analysis holds for general Gaussian-scale mixture prior distributions, including the HS prior as a special case of $\pi(\lambda) = C^+(\lambda|0,1)$. 
For the applicable distribution, however, the following conditions are imposed on $u(\lambda)\equiv-\frac{(\ln\pi(\lambda))'}{\lambda}$: 
\begin{enumerate}
    \item $u_0\equiv\lim_{\lambda\rightarrow0}u(\lambda)<\infty$,  
    \item $\lambda^4u(\lambda)=w_0^2$ has only one solution $\lambda^*$ with $\lambda>0$ for $\forall w_0\in\mathbb{R}$.
\end{enumerate}
One can easily verify that the half-Cauchy distribution satisfies these conditions; conversely, for $\pi(\lambda)$ satisfying these conditions, the recoverable phase can be derived by the same analysis. The properties of $\pi(\lambda)$ with respect to the phase transition and FPR are determined through the value of $\Delta\equiv\frac{\hat\chi}{u_0}$ at the fixed point; for a given $\alpha$ and $\rho$, a smaller value of $\Delta$ at the fixed point is associated with a larger the recoverable parameter region and lower FPR. This result is obtained when $\frac{\mathbb{E}_{w_0\sim\mathcal{N}(0,1)}\qty[u(\lambda^*)]}{u_0}$ is small. Finding a $\pi(\lambda)$ that is compatible with this property and sampling efficiency is a future challenge.

\begin{acknowledgments}
This work was supported by MEXT as the Program for Promoting Research on the Supercomputer Fugaku (DPMSD, Project ID: JPMXP1020200307). One of the authors, YN, was supported by the SPRING-GX program at the University of Tokyo.
\end{acknowledgments}

\appendix
\section{Derivation of RS free energy}
\label{sec:RS_free_energy}
This appendix describes in detail the derivation of the replica-symmetric free-energy density in Eq.~(\ref{eqn:RS_FED}). The replicated partition function to be evaluated is given as 
\begin{equation}
\begin{split}
\MoveEqLeft
    \mathbb{E}_{X,\bm w_0}\qty[Z^n(\beta;X,\bm w_0)]
    \\ &=\lim_{\gamma\rightarrow\infty}\mathbb{E}_{X,\bm w_0}\qty[\int\prod^n_{a=1}\dd{\bm w_a}e^{-\beta\sum^n_{a=1}\mathcal{L}(\bm w_0,\bm w_a)+\mathcal{H}(\bm w_a)}],
    \end{split}
\end{equation}
where the log-likelihood $\mathcal{L}$ is given by 
\begin{equation}
    \mathcal{L}(\bm w_0,\bm w_a) = \frac{\gamma}{2}\sum^M_{m=1}|\bm x_m^\top(\bm w_0-\bm w_a)|^2, 
\end{equation}
which imposes the constraint $X\bm w_0 = \bm X\bm w$ as $\gamma\rightarrow\infty$. The linear features $y_0$ and $y_a$ are defined as follows: 
\begin{equation}
\begin{split}
    y_0 &= \bm x^\top\bm w_0\\
    y_a &= \bm x^\top\bm w_a
\end{split}
\end{equation}
Noting that $\bm x$ contributes the free energy only through the linear features, $\mathcal{L}$ is simply denoted as 
\begin{equation}
    \mathcal{L}(y_0,y_a) = \frac{\gamma}{2}|y_0-y_a|^2. 
\end{equation}
The ``order parameters" are introduce by $\bm Q\in \mathbb{R}^{n\cross n}$ and $\bm m\in\mathbb{R}^n$ whose elements, $m_a$ and $Q_{ab}$ are defined using the covariances of the linear features as  
\begin{equation}
\begin{split}
    \mathbb{E}_{\bm x}[y_0y_a] &= \frac{1}{N}\bm w_0^\top\bm w_a \equiv m_a, \\
    \mathbb{E}_{\bm x}[y_ay_b] &= \frac{1}{N}\bm w_a^\top\bm w_b \equiv Q_{ab}, 
\end{split}
\end{equation}
respectively.
In addition to them, another parameter $\rho$ is defined by the covariance of $y_0$ as 
\begin{equation}
    \mathbb{E}_{\bm x}[y_0y_0] = \frac{1}{N}\bm w_0^\top\bm w_0 \equiv \rho. 
\end{equation}
This parameter $\rho$ also appears to be an order parameter, but it is trivially determined because of the quantity that is not relevant to the inference. 

By inserting these order parameters into the partition function as Lagrange multiplier 
with conjugate parameters $\tilde{\bm Q}\in \mathbb{R}^{n\cross n}$ and $\tilde{\bm m}\in\mathbb{R}^n$, the general formulation of replica free energy is obtained as \begin{align}
     &n\phi(\beta) =  \nonumber \\
     &\max_{\tilde{\bm Q},\tilde{\bm m},\bm Q,\bm m}\tilde\Psi(\tilde{\bm Q},\tilde{\bm m})+\frac{1}{2}\Tr(\tilde{\bm Q}\bm Q)+  \tilde{\bm m}^\top\bm m + \alpha\Psi(\bm Q,\bm m), 
    \label{equ:general_free_energy}
\end{align}
where
\begin{equation}
\begin{split}
    &\tilde\Psi(\tilde{\bm Q},\tilde{\bm m}) \\
    &= \mathbb{E}_{w_0}\qty[\int\dd{\bm w}\exp(-\frac{1}{2}\bm w^\top\tilde{\bm Q}\bm w - w_0\tilde{\bm m}^\top \bm w - \beta\mathcal{H}(\bm w))],
\end{split}
\end{equation}
\begin{equation}
    \Psi(\bm Q,\bm m) = \mathbb{E}_{y_0,\bm y}\qty[\exp(-\beta\sum^n_{a=1}\mathcal{L}(y_0,y_a))],
\end{equation}
and the expectations $\mathbb{E}_{w_0}[\cdots]$ and  $\mathbb{E}_{y_0,\bm y}[\cdots]$ means average over $w_0$ and $(y_0,\bm{y})$, respectively. 

Under the RS ansatz, the order parameters are reduced to six variables, $\chi$, $\hat{\chi}$, $q$, $\hat{Q}$, $m$ and $\hat{m}$ as follows: 
\begin{equation}
\begin{split}
    \bm Q &= \frac{\chi}{\beta} \bm I_n + q\bm 1_n,\\
    \tilde{\bm Q} &= \beta\hat Q\bm I_n - \beta^2\hat\chi\bm 1_n,\\
    m_a &= m, \\
    \tilde m_a &= -\beta \hat m, 
\end{split}
\label{eqn:RS_order_parameters}
\end{equation}
where $\bm I_n\in\mathbb{R}^{n\cross n}$ is the identity matrix and all entries of $\bm 1_n\in\mathbb{R}^{n\cross n}$ are $1$. 
This ansatz implies that $\bm y$ and $y_0$ are equivalent to the sum of the normal Gaussian variables $z_0$, $z$ and $z_a$, given by 
\begin{equation}
\begin{split}
    y_0 &= \sqrt{\rho-\frac{m^2}{q}}z_0 + \frac{m}{\sqrt{q}}z,\\
    y_a &= \sqrt{\frac{\chi}{\beta}}z_a + \sqrt{q}z. 
\end{split}
\end{equation}
By substituting these reduced RS order parameters in the general form of the free energy (\ref{equ:general_free_energy}), and insert Gaussian integration into the entropy term $\tilde{\bm \Psi}$ by using Hubbard-Stratonovich transformation, the RS free energy is obtained as 
\begin{align}
     \frac{1}{\beta}\phi(\beta;\alpha,\rho) &= \max_{\hat Q,\hat\chi,\hat m,q,\chi,m}\frac{1}{2} q\hat Q - \frac{1}{2}\chi\hat\chi - m\hat m\nonumber\\
     & + \alpha\Psi_{RS}(q,\chi,m;\rho) + \hat\Psi_{RS}(\hat Q,\hat\chi,\hat m), 
     \label{eqn:RS_FED_apendix}
\end{align}
where
\begin{equation}
    \Psi_{RS}(q,\chi,m;\rho) = -\frac{\rho-2m+q}{2\chi},
\end{equation}
and 
\begin{equation}\label{eqn:RS_entropy_term}
    \begin{split}
\MoveEqLeft
     \hat\Psi_{RS}(\hat Q,\hat\chi,\hat m) = \mathbb{E}_{w_0,\xi}\left[\frac{1}{\beta}\ln\int\dd{w} \right.\\
     & \left.\exp(\beta(-\frac{\hat Q}{2}|w|^2+(w_0\hat m + \sqrt{\hat\chi}\xi)w - \mathcal{H}(w)))\right].
    \end{split}
\end{equation}
$\hat\Psi_{RS}$ is usually called entropy term of RS free entropy.  Note that the expectation of any function of $w$ is obtained by integration with potential function in the entropy term, and this potential can be regarded as an effective posterior distribution. In the case of a hierarchical prior, $\mathcal{H}(w)$ is expressed as a Gaussian mixture distribution. Substituting Eq.(\ref{eqn:hierarchical_prior}), Eq.(\ref{eqn:hat_psi_rs}) is derived. 

\section{Perturbation at $\chi\rightarrow0$}
\label{Sec:B}
When CS is successful in the recoverable phase, the posterior distribution of $\bm w$ should converge to the true signal, and consequently $\hat Q$ should diverge to infinity in the recoverable phase. Then, the saddle point equations lead to $\chi\rightarrow0$.
At the limit of $\chi\rightarrow0$, the saddle point equations also show that $\hat Q=\hat m = \order{\chi^{-1}}$ and $\hat\chi=\order{1}$. 
 If there exists $\lambda^*>0$ such that partial derivative of the entropy term with respect to $\lambda$ equals to zero at $\lambda=\lambda^*$, Eq~(\ref{decide_variance}) leads to 
\begin{equation}\label{eqn:pdv_of_lambda}
    (\ln\pi(\lambda))'+\frac{\lambda(w_0\hat m+\sqrt{\hat\chi}\xi)^2}{(\hat Q\lambda^2+1)^2} = 0.
\end{equation}
Otherwise, under a plausible assumption, $\lambda^*=0$ is derived.

In the right-hand side of Eq.~(\ref{eqn:spe}), which is the saddle point equation for $\hat\chi$, the leading order of $\rho-2m+q$ is evaluated to 
\begin{equation}
\begin{split}
    &\rho-2m+q \\
    &=\rho + \mathbb{E}_{w_0,\xi}\qty[\left\langle\qty(\frac{w_0\hat m+\sqrt{\hat\chi}\xi}{\hat Q+\lambda^{-2}})^2-2w_0\qty(\frac{w_0\hat m+\sqrt{\hat\chi}\xi}{\hat Q+\lambda^{-2}})\right\rangle],\\
    &=\chi^2\mathbb{E}_{w_0,\xi}\qty[\left\langle\qty(\frac{\sqrt{\hat\chi}\lambda^2\xi-w_0}{\alpha \lambda^2+\chi })^2\right\rangle]. 
\end{split}
\end{equation}
Then, the leading term of $\lambda^{*2}$, which is the solution of Eq.~(\ref{eqn:pdv_of_lambda}), decides the recurrent equation of $\hat\chi$. By omitting negligible terms, the equation of $\hat\chi$ was derived as 
\begin{equation}
\begin{split}
    \hat\chi &= \alpha\left(\rho\frac{1}{\alpha^2}\qty[\left\langle\qty(\sqrt{\hat\chi}\xi-\frac{w_0}{\lambda^2})^2\right\rangle]_{\xi,w_0\sim\mathcal{N}(0,1)} \right.\\
    &~~~~\left.+ (1-\rho)\qty[\left\langle\qty(\frac{\sqrt{\hat\chi}\lambda^2\xi}{\alpha \lambda^2+\chi })^2\right\rangle]_{\xi\sim\mathcal{N}(0,1)}\right)\\
    &=\frac{1}{\alpha}\left(\rho\qty(\hat\chi+\qty[\left\langle\qty(\frac{w_0}{\lambda^2})^2\right\rangle]_{w_0\sim\mathcal{N}(0,1)} )\right.\\
    &\left.~~~~+ 2(1-\rho)\qty[\qty(\sqrt{\hat\chi}\xi-\sqrt{u_0})^2]_{\xi>\sqrt{\frac{u_0}{\hat\chi}}}\right)\\
    &=\frac{1}{\alpha}\left(\rho\qty(\hat\chi+\qty[ u(\lambda^*)]_{w_0\sim\mathcal{N}(0,1)} ) \right.\\
    &\left.~~+ 2(1-\rho)\qty((\hat\chi+u_0)\Phi\qty(\sqrt{\frac{u_0}{\hat\chi}})-\sqrt{\frac{u_0\hat\chi}{2\pi}}\exp(-\frac{u_0}{2\hat\chi}))\right)\\
\end{split}
\end{equation}
with some assumptions on $u(\lambda)$
\begin{equation}
    u(\lambda) \equiv -\frac{(\ln\pi(\lambda))'}{\lambda}>0,~u_0\equiv\lim_{\lambda\rightarrow0}u(\lambda) < \infty 
\end{equation}
and
\begin{equation}
    (\lambda^*)^4u(\lambda^*) = w_0^2,~\lambda^*<\infty.
\end{equation}
Here we use the notation $[\cdots]_{\xi>x}$ for $ \int_{x}D\xi\cdots$.
By normalizing $\hat\chi$ to $\Delta\equiv\frac{\hat\chi}{u_0}$, the recursive equation of $\Delta$ is derived as 
\begin{equation}
    \Delta = \frac{1}{\alpha}\qty(\rho\qty(\Delta+\frac{\qty[u(\lambda^*)]}{u_0}) + 2(1-\rho)\qty[\qty(\sqrt{\Delta}\xi-1)^2]_{\xi>\sqrt{\frac{1}{\Delta}}}). 
\end{equation}
The linear stability condition for the solution of this equation for  $\Delta$ yields Eq.~(\ref{linear_stability_condition}).

\section{Stability of RS ansatz}
\label{Sec:C}
To examine the stability of the RS solution, we construct a 1RSB solution with $n_1\cross n_1$ 1RSB blocks and define the order parameters at the 1RSB level as 
\begin{equation}
\begin{split}
    \bm Q &= \frac{\chi_0}{\beta}\bm I_n+\frac{\chi_1}{n_1\beta}\bm 1_{n_1}+q_0\bm 1_n,\\
    \tilde{\bm Q} &= \beta\hat Q\bm I_n-\beta^2\hat\chi_1\bm 1_{n_1}-\beta^2\hat\chi_0\bm 1_n,\\
    m_a &= m,\\
    \tilde m_a &= -\beta \hat m,
\end{split}
\end{equation}
where $\chi_1$ and $\hat{\chi_1}$ are small breaking parameters and $\bm 1_{n_1}\in\mathbb{R}^{n\cross n}$ is an $n\cross n$ block diagonal matrix consisting of $n_1\cross n_1$ blocks and all elements in the blocks are 1. 
The terms with $\chi_1$ and $\hat{\chi_1}$ represent the deviation from the RS order parameters of Eqs.~(\ref{eqn:RS_order_parameters}). 

If the RS ansatz gives an unstable solution, then the RS solution is expected to be destabilized for any $n_1$ and small breaking parameters $\chi_1$ and $\hat\chi_1$.
By substituting the 1RSB order parameters, the saddle point equations at $0<n_1\beta\ll 1$ are obtained by 
\begin{equation}
\begin{split}
    \hat m &= \frac{\alpha}{\chi_0+\chi_1},\\
    \hat Q &= \frac{\alpha}{\chi_0},\\
    \hat\chi_0 &= \alpha\frac{\rho-2m+q}{(\chi_1+\chi_0)^2},\\
    n_1\beta\hat\chi_1 &= \frac{\alpha\chi_1}{\chi_0(\chi_1+\chi_0)},\\
    \frac{1}{2}\qty(n_1\beta q_0 + \chi_1 + \chi_0) &= \frac{1}{2\sqrt{\hat\chi_1}}\mathbb{E}_{\xi_0,\xi_1,w_0}\qty[\xi_1w^*],\\
    \frac{1}{2}\qty(\chi_1+\chi_0) &= \frac{1}{2\sqrt{\hat\chi_0}}\mathbb{E}_{\xi_0,\xi_1,w_0}\qty[\xi_0w^*],\\
    \frac{1}{2}\qty(\frac{\chi_0}{\beta}+\frac{\chi_1}{n_1\beta}+q_0) &= \mathbb{E}_{\xi_0,\xi_1,w_0}\qty[\frac{|w^*|^2}{2}],\\
    m &= \mathbb{E}_{\xi_0,\xi_1,w_0}\qty[w_0w^*]
\end{split}
\end{equation}
where $w^*$ is determined by the saddle point condition; 
\begin{equation}
    w^* = \argmax_w -\frac{\hat Q}{2}w^2 + (\sqrt{\hat\chi_1}\xi_1 + \sqrt{\hat\chi_0}\xi_0 + w_0\hat m)w - \mathcal{H}(w).
\end{equation}
For $\chi_1,\hat\chi_1 \ll 1$, if the order parameters other than $\chi_1$ and $\hat{\chi_1}$ have the same values as the RS solution, the equation to be satisfied by $\hat\chi_1$ is obtained as 
\begin{equation}
    \hat\chi_1 = \frac{\rho +2(1-\rho)\Phi(\sqrt{\frac{2}{\hat\chi_0}})}{\alpha}\hat\chi_1. 
\end{equation}
Considered as a recursive equation, the condition for $\hat\chi_1$ to converge to 0 is derived as $\frac{\rho+2(1-\rho)\Phi(\sqrt{\frac{2}{\hat\chi_0}})}{\alpha}<1$. 
Since this condition coincides with the linear stability for $\hat\chi$ in RS solution, Eq.~(\ref{linear_stability_condition}),  we conclude that the RS analysis is stable in the recoverbility phase.



\begin{thebibliography}{0}%
\makeatletter
\providecommand \@ifxundefined [1]{%
 \@ifx{#1\undefined}
}%
\providecommand \@ifnum [1]{%
 \ifnum #1\expandafter \@firstoftwo
 \else \expandafter \@secondoftwo
 \fi
}%
\providecommand \@ifx [1]{%
 \ifx #1\expandafter \@firstoftwo
 \else \expandafter \@secondoftwo
 \fi
}%
\providecommand \natexlab [1]{#1}%
\providecommand \enquote  [1]{``#1''}%
\providecommand \bibnamefont  [1]{#1}%
\providecommand \bibfnamefont [1]{#1}%
\providecommand \citenamefont [1]{#1}%
\providecommand \href@noop [0]{\@secondoftwo}%
\providecommand \href [0]{\begingroup \@sanitize@url \@href}%
\providecommand \@href[1]{\@@startlink{#1}\@@href}%
\providecommand \@@href[1]{\endgroup#1\@@endlink}%
\providecommand \@sanitize@url [0]{\catcode `\\12\catcode `\$12\catcode
  `\&12\catcode `\#12\catcode `\^12\catcode `\_12\catcode `\%12\relax}%
\providecommand \@@startlink[1]{}%
\providecommand \@@endlink[0]{}%
\providecommand \url  [0]{\begingroup\@sanitize@url \@url }%
\providecommand \@url [1]{\endgroup\@href {#1}{\urlprefix }}%
\providecommand \urlprefix  [0]{URL }%
\providecommand \Eprint [0]{\href }%
\providecommand \doibase [0]{https://doi.org/}%
\providecommand \selectlanguage [0]{\@gobble}%
\providecommand \bibinfo  [0]{\@secondoftwo}%
\providecommand \bibfield  [0]{\@secondoftwo}%
\providecommand \translation [1]{[#1]}%
\providecommand \BibitemOpen [0]{}%
\providecommand \bibitemStop [0]{}%
\providecommand \bibitemNoStop [0]{.\EOS\space}%
\providecommand \EOS [0]{\spacefactor3000\relax}%
\providecommand \BibitemShut  [1]{\csname bibitem#1\endcsname}%
\let\auto@bib@innerbib\@empty
\end{thebibliography}%


\begin{thebibliography}{26}
\expandafter\ifx\csname natexlab\endcsname\relax\def\natexlab#1{#1}\fi
\expandafter\ifx\csname bibnamefont\endcsname\relax
  \def\bibnamefont#1{#1}\fi
\expandafter\ifx\csname bibfnamefont\endcsname\relax
  \def\bibfnamefont#1{#1}\fi
\expandafter\ifx\csname citenamefont\endcsname\relax
  \def\citenamefont#1{#1}\fi
\expandafter\ifx\csname url\endcsname\relax
  \def\url#1{\texttt{#1}}\fi
\expandafter\ifx\csname urlprefix\endcsname\relax\def\urlprefix{URL }\fi
\providecommand{\bibinfo}[2]{#2}
\providecommand{\eprint}[2][]{\url{#2}}

\bibitem[{\citenamefont{Donoho}(2006)}]{donoho2006compressed}
\bibinfo{author}{\bibfnamefont{D.~L.} \bibnamefont{Donoho}},
  \bibinfo{journal}{IEEE Trans. Inf. Theory}
  \textbf{\bibinfo{volume}{52}}, \bibinfo{pages}{1289} (\bibinfo{year}{2006}).

\bibitem[{\citenamefont{Cand{\`e}s and Wakin}(2008)}]{candes2008introduction}
\bibinfo{author}{\bibfnamefont{E.~J.} \bibnamefont{Cand{\`e}s}}
  \bibnamefont{and} \bibinfo{author}{\bibfnamefont{M.~B.} \bibnamefont{Wakin}},
  \bibinfo{journal}{IEEE Signal Process. Mag.}
  \textbf{\bibinfo{volume}{25}}, \bibinfo{pages}{21} (\bibinfo{year}{2008}).

\bibitem[{\citenamefont{Donoho and Tanner}(2009)}]{donoho2009observed}
\bibinfo{author}{\bibfnamefont{D.}~\bibnamefont{Donoho}} \bibnamefont{and}
  \bibinfo{author}{\bibfnamefont{J.}~\bibnamefont{Tanner}},
  \bibinfo{journal}{Philos. Trans. of the R. Soc., A}
  \textbf{\bibinfo{volume}{367}}, \bibinfo{pages}{4273} (\bibinfo{year}{2009}).

\bibitem[{\citenamefont{Tibshirani}(1996)}]{tibshirani1996regression}
\bibinfo{author}{\bibfnamefont{R.}~\bibnamefont{Tibshirani}},
  \bibinfo{journal}{J. R. Stat. Soc.: Ser. B} \textbf{\bibinfo{volume}{58}}, \bibinfo{pages}{267}
  (\bibinfo{year}{1996}).

\bibitem[{\citenamefont{Tibshirani}(1997)}]{tibshirani1997lasso}
\bibinfo{author}{\bibfnamefont{R.}~\bibnamefont{Tibshirani}},
  \bibinfo{journal}{Stat. Med.} \textbf{\bibinfo{volume}{16}},
  \bibinfo{pages}{385} (\bibinfo{year}{1997}).

\bibitem[{\citenamefont{Chen et~al.}(2001)\citenamefont{Chen, Donoho, and
  Saunders}}]{chen2001atomic}
\bibinfo{author}{\bibfnamefont{S.~S.} \bibnamefont{Chen}},
  \bibinfo{author}{\bibfnamefont{D.~L.} \bibnamefont{Donoho}},
  \bibnamefont{and} \bibinfo{author}{\bibfnamefont{M.~A.}
  \bibnamefont{Saunders}}, \bibinfo{journal}{SIAM Rev.}
  \textbf{\bibinfo{volume}{43}}, \bibinfo{pages}{129} (\bibinfo{year}{2001}).

\bibitem[{\citenamefont{Beck and Teboulle}(2009)}]{beck2009fast}
\bibinfo{author}{\bibfnamefont{A.}~\bibnamefont{Beck}} \bibnamefont{and}
  \bibinfo{author}{\bibfnamefont{M.}~\bibnamefont{Teboulle}},
  \bibinfo{journal}{SIAM J. Imaging Sci.}
  \textbf{\bibinfo{volume}{2}}, \bibinfo{pages}{183} (\bibinfo{year}{2009}).

\bibitem[{\citenamefont{Park and Casella}(2008)}]{park2008bayesian}
\bibinfo{author}{\bibfnamefont{T.}~\bibnamefont{Park}} \bibnamefont{and}
  \bibinfo{author}{\bibfnamefont{G.}~\bibnamefont{Casella}},
  \bibinfo{journal}{J. Am. Stat. Assoc.}
  \textbf{\bibinfo{volume}{103}}, \bibinfo{pages}{681} (\bibinfo{year}{2008}).

\bibitem[{\citenamefont{Carvalho et~al.}(2009)\citenamefont{Carvalho, Polson,
  and Scott}}]{carvalho2009handling}
\bibinfo{author}{\bibfnamefont{C.~M.} \bibnamefont{Carvalho}},
  \bibinfo{author}{\bibfnamefont{N.~G.} \bibnamefont{Polson}},
  \bibnamefont{and} \bibinfo{author}{\bibfnamefont{J.~G.} \bibnamefont{Scott}},
  in \emph{\bibinfo{booktitle}{Artificial Intelligence and Statistics}}
  (\bibinfo{organization}{PMLR}, \bibinfo{year}{2009}), pp.
  \bibinfo{pages}{73--80}.

\bibitem[{\citenamefont{Carvalho et~al.}(2010)\citenamefont{Carvalho, Polson,
  and Scott}}]{carvalho2010horseshoe}
\bibinfo{author}{\bibfnamefont{C.~M.} \bibnamefont{Carvalho}},
  \bibinfo{author}{\bibfnamefont{N.~G.} \bibnamefont{Polson}},
  \bibnamefont{and} \bibinfo{author}{\bibfnamefont{J.~G.} \bibnamefont{Scott}},
  \bibinfo{journal}{Biometrika} \textbf{\bibinfo{volume}{97}},
  \bibinfo{pages}{465} (\bibinfo{year}{2010}).

\bibitem[{\citenamefont{Bhadra et~al.}(2019)\citenamefont{Bhadra, Datta,
  Polson, and Willard}}]{bhadra2019lasso}
\bibinfo{author}{\bibfnamefont{A.}~\bibnamefont{Bhadra}},
  \bibinfo{author}{\bibfnamefont{J.}~\bibnamefont{Datta}},
  \bibinfo{author}{\bibfnamefont{N.~G.} \bibnamefont{Polson}},
  \bibnamefont{and} \bibinfo{author}{\bibfnamefont{B.}~\bibnamefont{Willard}},
  \bibinfo{journal}{Stat. Sci.} \textbf{\bibinfo{volume}{34}},
  \bibinfo{pages}{405} (\bibinfo{year}{2019}).

\bibitem[{\citenamefont{Van Der~Pas et~al.}(2014)\citenamefont{Van Der~Pas,
  Kleijn, and Van Der~Vaart}}]{van2014horseshoe}
\bibinfo{author}{\bibfnamefont{S.~L.} \bibnamefont{Van Der~Pas}},
  \bibinfo{author}{\bibfnamefont{B.~J.} \bibnamefont{Kleijn}},
  \bibnamefont{and} \bibinfo{author}{\bibfnamefont{A.~W.} \bibnamefont{Van
  Der~Vaart}}, \bibinfo{journal}{Electron. J. Stat.}
  \textbf{\bibinfo{volume}{8}}, \bibinfo{pages}{2585} (\bibinfo{year}{2014}).

\bibitem[{\citenamefont{Gelman}(2006)}]{gelman2006prior}
\bibinfo{author}{\bibfnamefont{A.}~\bibnamefont{Gelman}},
  \bibinfo{journal}{Bayesian Anal.} \textbf{\bibinfo{volume}{1}},
  \bibinfo{pages}{515} (\bibinfo{year}{2006}).

\bibitem[{\citenamefont{Polson and Scott}(2012)}]{polson2012half}
\bibinfo{author}{\bibfnamefont{N.~G.} \bibnamefont{Polson}} \bibnamefont{and}
  \bibinfo{author}{\bibfnamefont{J.~G.} \bibnamefont{Scott}},
  \bibinfo{journal}{Bayesian Anal.} \textbf{\bibinfo{volume}{7}},
  \bibinfo{pages}{887} (\bibinfo{year}{2012}).

\bibitem[{\citenamefont{Van Der~Pas et~al.}(2016)\citenamefont{Van Der~Pas,
  Salomond, and Schmidt-Hieber}}]{van2016conditions}
\bibinfo{author}{\bibfnamefont{S.}~\bibnamefont{Van Der~Pas}},
  \bibinfo{author}{\bibfnamefont{J.-B.} \bibnamefont{Salomond}},
  \bibnamefont{and}
  \bibinfo{author}{\bibfnamefont{J.}~\bibnamefont{Schmidt-Hieber}},
  \bibinfo{journal}{Electron. J. Stat.}
  \textbf{\bibinfo{volume}{10}}, \bibinfo{pages}{976} (\bibinfo{year}{2016}).

\bibitem[{\citenamefont{Ghosh et~al.}(2016)\citenamefont{Ghosh, Tang, Ghosh,
  and Chakrabarti}}]{ghosh2016asymptotic}
\bibinfo{author}{\bibfnamefont{P.}~\bibnamefont{Ghosh}},
  \bibinfo{author}{\bibfnamefont{X.}~\bibnamefont{Tang}},
  \bibinfo{author}{\bibfnamefont{M.}~\bibnamefont{Ghosh}}, \bibnamefont{and}
  \bibinfo{author}{\bibfnamefont{A.}~\bibnamefont{Chakrabarti}},
  \bibinfo{journal}{Bayesian Anal.} \textbf{\bibinfo{volume}{11}},
  \bibinfo{pages}{753} (\bibinfo{year}{2016}).

\bibitem[{\citenamefont{Makalic and Schmidt}(2015)}]{makalic2015simple}
\bibinfo{author}{\bibfnamefont{E.}~\bibnamefont{Makalic}} \bibnamefont{and}
  \bibinfo{author}{\bibfnamefont{D.~F.} \bibnamefont{Schmidt}},
  \bibinfo{journal}{IEEE Signal Process. Lett.}
  \textbf{\bibinfo{volume}{23}}, \bibinfo{pages}{179} (\bibinfo{year}{2015}).

\bibitem[{\citenamefont{Bhattacharya et~al.}(2016)\citenamefont{Bhattacharya,
  Chakraborty, and Mallick}}]{bhattacharya2016fast}
\bibinfo{author}{\bibfnamefont{A.}~\bibnamefont{Bhattacharya}},
  \bibinfo{author}{\bibfnamefont{A.}~\bibnamefont{Chakraborty}},
  \bibnamefont{and} \bibinfo{author}{\bibfnamefont{B.~K.}
  \bibnamefont{Mallick}}, \bibinfo{journal}{Biometrika} \textbf{\bibinfo{volume}{103}}, \bibinfo{pages}{985} (\bibinfo{year}{2016}).

\bibitem[{\citenamefont{Kabashima et~al.}(2009)\citenamefont{Kabashima,
  Wadayama, and Tanaka}}]{kabashima2009typical}
\bibinfo{author}{\bibfnamefont{Y.}~\bibnamefont{Kabashima}},
  \bibinfo{author}{\bibfnamefont{T.}~\bibnamefont{Wadayama}}, \bibnamefont{and}
  \bibinfo{author}{\bibfnamefont{T.}~\bibnamefont{Tanaka}},
  \bibinfo{journal}{J. Stat. Mech.: Theory Exp.}
  \textbf{\bibinfo{volume}{2009}}, \bibinfo{pages}{L09003}
  (\bibinfo{year}{2009}).

\bibitem[{\citenamefont{Ganguli and
  Sompolinsky}(2010)}]{ganguli2010statistical}
\bibinfo{author}{\bibfnamefont{S.}~\bibnamefont{Ganguli}} \bibnamefont{and}
  \bibinfo{author}{\bibfnamefont{H.}~\bibnamefont{Sompolinsky}},
  \bibinfo{journal}{Phys. rev. Lett.} \textbf{\bibinfo{volume}{104}},
  \bibinfo{pages}{188701} (\bibinfo{year}{2010}).

\bibitem[{\citenamefont{Donoho et~al.}(2009)\citenamefont{Donoho, Maleki, and
  Montanari}}]{donoho2009message}
\bibinfo{author}{\bibfnamefont{D.~L.} \bibnamefont{Donoho}},
  \bibinfo{author}{\bibfnamefont{A.}~\bibnamefont{Maleki}}, \bibnamefont{and}
  \bibinfo{author}{\bibfnamefont{A.}~\bibnamefont{Montanari}},
  \bibinfo{journal}{Proc. Natl. Acad. Sci. U.S.A.}
  \textbf{\bibinfo{volume}{106}}, \bibinfo{pages}{18914}
  (\bibinfo{year}{2009}).

\bibitem[{\citenamefont{Donoho et~al.}(2010)\citenamefont{Donoho, Maleki, and
  Montanari}}]{donoho2010message_2}
\bibinfo{author}{\bibfnamefont{D.~L.} \bibnamefont{Donoho}},
  \bibinfo{author}{\bibfnamefont{A.}~\bibnamefont{Maleki}}, \bibnamefont{and}
  \bibinfo{author}{\bibfnamefont{A.}~\bibnamefont{Montanari}}, in
  \emph{\bibinfo{booktitle}{2010 IEEE Information Theory Workshop on
  Information Theory (ITW 2010, Cairo)}} (\bibinfo{organization}{IEEE},
  \bibinfo{year}{2010}), pp. \bibinfo{pages}{1--5}.

\bibitem[{\citenamefont{Caltagirone et~al.}(2014)\citenamefont{Caltagirone,
  Zdeborov{\'a}, and Krzakala}}]{caltagirone2014convergence}
\bibinfo{author}{\bibfnamefont{F.}~\bibnamefont{Caltagirone}},
  \bibinfo{author}{\bibfnamefont{L.}~\bibnamefont{Zdeborov{\'a}}},
  \bibnamefont{and} \bibinfo{author}{\bibfnamefont{F.}~\bibnamefont{Krzakala}},
  in \emph{\bibinfo{booktitle}{2014 IEEE International Symposium on Information
  Theory}} (\bibinfo{organization}{IEEE}, \bibinfo{year}{2014}), pp.
  \bibinfo{pages}{1812--1816}.

\bibitem[{\citenamefont{Krzakala
  et~al.}(2012{\natexlab{a}})\citenamefont{Krzakala, M{\'e}zard, Sausset, Sun,
  and Zdeborov{\'a}}}]{krzakala2012statistical}
\bibinfo{author}{\bibfnamefont{F.}~\bibnamefont{Krzakala}},
  \bibinfo{author}{\bibfnamefont{M.}~\bibnamefont{M{\'e}zard}},
  \bibinfo{author}{\bibfnamefont{F.}~\bibnamefont{Sausset}},
  \bibinfo{author}{\bibfnamefont{Y.}~\bibnamefont{Sun}}, \bibnamefont{and}
  \bibinfo{author}{\bibfnamefont{L.}~\bibnamefont{Zdeborov{\'a}}},
  \bibinfo{journal}{Phys. Rev. X} \textbf{\bibinfo{volume}{2}},
  \bibinfo{pages}{021005} (\bibinfo{year}{2012}{\natexlab{a}}).

\bibitem[{\citenamefont{Krzakala
  et~al.}(2012{\natexlab{b}})\citenamefont{Krzakala, M{\'e}zard, Sausset, Sun,
  and Zdeborov{\'a}}}]{krzakala2012probabilistic}
\bibinfo{author}{\bibfnamefont{F.}~\bibnamefont{Krzakala}},
  \bibinfo{author}{\bibfnamefont{M.}~\bibnamefont{M{\'e}zard}},
  \bibinfo{author}{\bibfnamefont{F.}~\bibnamefont{Sausset}},
  \bibinfo{author}{\bibfnamefont{Y.}~\bibnamefont{Sun}}, \bibnamefont{and}
  \bibinfo{author}{\bibfnamefont{L.}~\bibnamefont{Zdeborov{\'a}}},
  \bibinfo{journal}{J. Stat. Mech.: Theory Exp.}
  \textbf{\bibinfo{volume}{2012}}, \bibinfo{pages}{P08009}
  (\bibinfo{year}{2012}{\natexlab{b}}).

\bibitem[{\citenamefont{Talagrand et~al.}(2003)}]{talagrand2003spin}
\bibinfo{author}{\bibfnamefont{M.}~\bibnamefont{Talagrand}}
  \bibnamefont{et~al.}, \bibinfo{title}{A Series of Modern Surveys in Mathematics}, vol.~\bibinfo{volume}{46}
  (\bibinfo{publisher}{Springer Science \& Business Media},
  \bibinfo{year}{2003}).

\end{thebibliography}
\end{document}